# Finite-time synchronization between two different chaotic systems with uncertainties

Meili Lin, Zhengzhong Yuan, Jianping Cai*

Department of Mathematics and Information Science, Zhangzhou Normal University, Zhangzhou 363000, China

*Corresponding author, Email: mathcai@hotmail.com, Tel: 86-596-2527285, Fax: 86-596-2527012

**Abstract** A new method of virtual unknown parameter is proposed to synchronize two different systems with unknown parameters and disturbance in finite time. Virtual unknown parameters are introduced in order to avoid the unknown parameters from appearing in the controllers and parameters update laws when the adaptive control method is applied. A single virtual unknown parameter is used in the design of adaptive controllers and parameters update laws if the Lipschitz constant on the nonlinear function can be found, while multiple virtual unknown parameters are adopted if the Lipschitz constant cannot be determined. Numerical simulations show that the present method does make the two different chaotic systems synchronize in finite time.

## 1. Introduction

Synchronization in chaotic dynamic systems has attracted increasing attention of scientists from various research fields for its advantages in practical applications [1]. A wide variety of methods have been proposed for synchronization of chaotic systems, including linear feedback control [2,3], sliding mode control [4,5], adaptive control [6,7] and so on. Most of the methods mentioned above are used to guarantee the asymptotic stability of chaotic systems. In other words, the convergence of the synchronization procedure is asymptotic with infinite settling time.

In many practical applications, one of the main concerns is that the behavior of system is stable over a finite-time interval. To achieve faster convergence in control systems, finite-time control is a very useful technique. The finite-time control techniques have demonstrated better robustness and disturbance rejection properties [8-10]. For the nonlinear system of the form [11]

$$\begin{cases} \dot{x}_j = x_{j+1}^{m_j} + \phi_j(x_1, \cdots x_j) & (j = 1, \cdots n-1) \\ \dot{x}_n = u + \phi_n(x_1, \cdots x_n) \end{cases}$$

where $m_j$ are odd integers, output feedback controllers can be explicitly constructed to globally stabilize the system in finite time [11-15]. Finite-time observers for nonlinear systems that can be put into a linear canonical form up to the output injection have been proposed [16-18]. Whereas, linearization method just ensure the local stability. Methods in [11-15] are suitable only for the special systems mentioned above, and cannot be applied directly to synchronization for general chaotic systems, such as Lorenz system and Chen system.

So far, adopting the concept of convergence in finite time [19], finite-time sliding-mode controllers and feedback controllers have become two effective methods in finite-time synchronization. The state feedback control law involving the term $-k_1 sign(x)|x|^{\alpha_1} - k_2 sign(v)|v|^{\alpha_2}$ is developed to solve the finite-time synchronization problem of two different second-order chaotic systems without uncertainty [8]. However the controller in [8] contains all the information appeared in the error dynamical system or the drive-response systems, which is hard to be realized in practical application.

Nevertheless uncertainties do exist in any real world systems. The finite-time control problems for linear systems subject to time-varying parametric uncertainties and to exogenous constant disturbances is considered[20]. A sliding-mode controller is designed to drive a class of chaotic systems with nonlinear inputs to track a smooth target trajectory in finite time[21]. The nonsingular terminal sliding mode control is used to stabilize the chaotic systems with uncertain parameters or disturbances in finite time [22]. A controller containing the term $e^{\beta}$ with $\beta = q/p < 1$ is proposed to realize the finite-time synchronization of two identical unified chaotic systems with uncertain parameters[23]. In view of a control Lyapunov function, controllers including the term of Lie derivative $(L_B V)^T L_f V$ are designed to cope with the chaos synchronization for the coupled identical united chaotic system with uncertain parameters[24].

In this paper, we introduce a virtual unknown parameter method to realize the finite-time synchronization between two different systems with uncertainties. In order to avoid the unknown parameters from appearing in controllers and parameters update laws, virtual unknown parameters are introduced. With the help of a virtual parameter, an adaptive controller and the corresponding parameters update laws are therefore designed to synchronize two coupled chaotic systems with uncertainties in finite time when the Lipschitz constant on the nonlinear function can be found. Sometimes it is hard to determine the Lipschitz constant. In order to overcome this difficulty, we introduce multiple virtual unknown parameters to develop techniques for obtaining proper adaptive controllers. Some typical chaotic systems, Rössler system, Chua's circuit and Lorenz system, are chosen as illustrative examples to demonstrate the effectiveness of the proposed method.

## 2. Problem statement

Consider a class of chaotic systems with unknown parameters and disturbance in the form of

$$\dot{x} = f(x) + F(x)\alpha + d(x,t), \tag{1}$$

where $x \in R^n$ is the state vector, $\alpha \in R^m$ is an unknown parameter vector, $f(x) \in R^{n\times 1}$ and $F(x) \in R^{n\times m}$ are nonlinear matrix functions, $d(x,t) \in R^n$ is the external disturbance. Let $\Omega_x \subset R^n$ be a bounded region containing the whole attractor of drive system (1) such that no trajectory of system (1) ever leaves it. This assumption is simply based on the bounded property of chaotic attractor. Also, let $M \subset R^m$ be the set of parameter under which system (1) is in a chaotic state.

Introducing a virtual unknown parameter $\alpha_{m+1}$, Eq.(1) can be transformed into

$$\dot{x} = \bar{f}(x) + \bar{F}(x)\bar{\alpha} + d(x,t) \tag{2}$$

where $\bar{f}(x) \in R^{n\times 1}$ and $\bar{F}(x) \in R^{n\times(m+1)}$ are nonlinear matrix functions , $\bar{\alpha}^T = (\alpha^T, \alpha_{m+1}) \in R^{m+1}$ is a new parameter vector, and the term in $\bar{F}(x)$ corresponding to virtual unknown parameter $\alpha_{m+1}$ is called virtual term.

Many chaotic systems can be transformed into the form of system (1) and (2). Let us take Lorenz system as an example,

$$\begin{pmatrix} \dot{x}_1 \\ \dot{x}_2 \\ \dot{x}_3 \end{pmatrix} = \begin{pmatrix} 0 \\ -x_2 - x_1 x_3 \\ x_1 x_2 \end{pmatrix} + \begin{pmatrix} x_2 - x_1 & 0 & 0 \\ 0 & x_1 & 0 \\ 0 & 0 & x_3 \end{pmatrix} \begin{pmatrix} \alpha_1 \\ \alpha_2 \\ \alpha_3 \end{pmatrix} = f(x) + F(x)\alpha$$

$$= \begin{pmatrix} 0 \\ -x_1 x_3 \\ x_1 x_2 \end{pmatrix} + \begin{pmatrix} x_2 - x_1 & 0 & 0 & 0 \\ 0 & x_1 & 0 & -x_2 \\ 0 & 0 & x_3 & 0 \end{pmatrix} \begin{pmatrix} \alpha_1 \\ \alpha_2 \\ \alpha_3 \\ \alpha_4 \end{pmatrix} = \bar{f}(x) + \bar{F}(x)\bar{\alpha}.$$

where $\alpha_1, \alpha_2, \alpha_3$ are unknown parameters, $\alpha_4 = 1$ is a virtual unknown parameter, and $-x_2$ is the virtual term corresponding to virtual unknown parameter $\alpha_4$.

**Remark 1.** Each component, linear or nonlinear, of $f(x)$ can be elected as a virtual term in $\bar{F}(x)$ corresponding to parameter $\alpha_{m+1}$, where $\alpha_{m+1} \neq 0$.

## 3. Finite-time chaos synchronization

### 3.1 A single virtual unknown parameter in response system

Let Eq. (2) be the drive system. The corresponding response system is another different system defined as below,

$$\dot{y} = g(y) + G(y)\beta + u(t,x) \tag{3}$$

which can be transformed into

$$\dot{y} = \bar{g}(y) + \bar{G}(y)\bar{\beta} + u(t,x) \tag{4}$$

where $y \in R^n$ is the state vector, $g(y) \in R^{n\times 1}$, $\bar{g}(y) \in R^{n\times 1}$, $G(y) \in R^{n\times l}$ and $\bar{G}(y) \in R^{n\times(l+1)}$ are nonlinear matrix functions. $\bar{\beta}^T = (\beta^T, \beta_{l+1}) \in R^{l+1}$ is a new parameter vector, in which $\beta \in R^l$ is an unknown parameter vector of the initial system and $\beta_{l+1}$ is a virtual unknown parameter. $u(t,x)$ is a controller. Let $\Omega_y \subset R^n$ be a bounded region containing the whole attractor of response system (4) with $u(t,x) = 0$.

The object of this paper is that the coupled system (2) and (4) can be synchronized in finite time by designing an effective controller $u(t,x)$ and adaptive laws. In order to do so, some hypothesizes are made as follows,

**H1.** The nonlinear functions $\bar{f}(\cdot)$ and $\bar{g}(\cdot)$ are smooth and continuous on a bounded closed region $\Omega$ containing both $\Omega_x$ and $\Omega_y$. Thus there exists a constant $M$ such that

$$\left\|\bar{f}(x) - \bar{g}(x)\right\| \le M.$$

**H2.** The nonlinear matrix function $\bar{g}(\cdot)$ satisfies Lipschitz condition, namely

$$\left\|\bar{g}(x) - \bar{g}(y)\right\| \le L_{\bar{g}} \left\|x - y\right\|$$

where $L_{\bar{g}}$ is an appropriate positive constant. In this paper, $\|\bullet\|$ denotes matrix or vector norms, defined as $\|A\| = (\sum_{j=1}^{m}\sum_{i=1}^{n} a_{ij}^2)^{1/2}$ for the matrix $A = (a_{ij})_{n\times m}$.

Fortunately, many chaotic systems transformed into the form of Eq.(2) and (4) satisfy assumptions H1 and H2, such as Lorenz system, Rössler system, Genesio-Tesi system.

**H3.** The uncertain parameters $\alpha, \beta$ and disturbance $d(x,t)$ are all norm bounded,

$$\|\alpha\| \le \delta_\alpha, \|\beta\| \le \delta_\beta, \|d(x,t)\| \le \delta_d$$

where $\delta_\alpha, \delta_\beta, \delta_d$ are positive constants. Then there exist two positive constants $\bar{\delta}_\alpha$ and $\bar{\delta}_\beta$, such that

$$\|\bar{\alpha}\| = \bar{\delta}_\alpha \le \delta_\alpha + |\alpha_{m+1}|, \|\bar{\beta}\| = \bar{\delta}_\beta \le \delta_\beta + |\beta_{l+1}| .$$

Introduce the following notations,

$$\Delta_\alpha = (\bar{\delta}_\alpha, \bar{\delta}_\alpha \cdots \bar{\delta}_\alpha)^T, \quad \Delta_\beta = (\bar{\delta}_\beta, \bar{\delta}_\beta \cdots \bar{\delta}_\beta)^T,$$

which are $m+1$-dimensional and $l+1$-dimensional constant vectors, respectively.

Define the error variable as $e = x - y$. Subtracting (4) from (2) yields the synchronization error dynamics as

$$\dot{e} = f(x) - g(y) + F(x)\bar{\alpha} - G(y)\bar{\beta} + d(x,t) \tag{5}$$

So the problem to realize synchronization between systems (2) and (4) in finite time is transformed into the finite time stability of error dynamics (5). Hence the definition of finite-time synchronization and some necessary lemmas are introduced in the follows.

**Definition** [23] Consider the following two chaotic systems:

$$\begin{aligned} \dot{x}_m &= f(x_m), \\ \dot{x}_s &= h(x_m, x_s), \end{aligned} \tag{6}$$

where $x_m, x_s$ are two $n$-dimensional vectors. The subscripts '$m$' and '$n$' stand for the master and slave systems, respectively. $f: R^n \to R^n$ and $h: R^n \to R^n$ are vector-valued functions. If there exists a constant $T > 0$, such that

$$\lim_{t \to T} \|x_m - x_s\| = 0$$

and $\|x_m - x_s\| \equiv 0$, if $t \ge T$, then synchronization of the system (6) is achieved in a finite-time.

**Lemma 1**[23] Assume that a continuous, positive-definite function $V(t)$ satisfies the following differential inequality,

$$\dot{V}(t) \le -\ c V(\ ,\quad \forall t > 0, V(t_0) \ge 0,$$

where $c > 0, 0 < \eta < 1$ are all constants. Then, for any given $t_0$, $V(t)$ satisfies the following inequality,

$$V^{1-\eta}(t) \le V^{1-\eta}(t_0) - c(1-\eta)(t-t_0), t_0 \le t \le t_1,$$

and $V(t) \equiv 0$ for $t > t_1$ with $t_1 = t_0 + \dfrac{V^{1-\eta}(t_0)}{c(1-\eta)}$.

**Lemma 2**[11] For any given $a, b \in R, 0 < q \le 1$, the following inequality holds,

$$\left(|a| + |b|\right)^q \le |a|^q + |b|^q .$$

Now we can establish the following result.

**Theorem 1.** The coupled system (2) and (4) can be synchronized in finite time if the hypotheses H1~H3 hold and the following conditions $(\mathrm{I})-(\Pi)$ are satisfied,

(I) The controller is chosen as

$$\begin{aligned} u(x,t) &= k_1 e + (k_2+1)\frac{e}{\|e\|} + \bar{F}(x)\hat{\alpha} - \bar{G}(y)\hat{\beta} \\ &+2\frac{\|\Delta_\alpha\|^2 + \|\Delta_\alpha\|\cdot\|\hat{\alpha}\|}{|\alpha_{m+1} - \hat{\alpha}_{m+1}|}\cdot\frac{e}{\|e\|^2} + 2\frac{\|\Delta_\beta\|^2 + \|\Delta_\beta\|\cdot\|\hat{\beta}\|}{|\beta_{l+1} - \hat{\beta}_{l+1}|}\cdot\frac{e}{\|e\|^2} \end{aligned} \quad (7)$$

where $k_1 \ge L_{\bar{g}}, k_2 \ge \delta_d + M$, $\hat{\alpha}$ and $\hat{\beta}$ are two vector parameters.

$(\Pi)$ The update laws of the parameters $\hat{\alpha}$ and $\hat{\beta}$ are

$$\dot{\hat{\alpha}} = \bar{F}(x)^T e + \frac{\Delta_\alpha - \hat{\alpha}}{|\alpha_{m+1} - \hat{\alpha}_{m+1}|}, \quad (8)$$

$$\dot{\hat{\beta}} = -\bar{G}(y)^T e + \frac{\Delta_\beta - \hat{\beta}}{|\beta_{l+1} - \hat{\beta}_{l+1}|} \quad (9)$$

where $\hat{\alpha}_{m+1}$ and $\hat{\beta}_{l+1}$ represent the $(m+1)-th$ and $(l+1)-th$ component of vectors $\hat{\alpha}$ and $\hat{\beta}$, respectively.

For a fixed $t_0$, the finite time $t_1$ is determined by

$$t_1 = t_0 + 2^{\frac{1}{2}} V^{\frac{1}{2}}(t_0) \quad . \tag{10}$$

where $V$ is a positive-definite function satisfying Lemma 1.

**Proof**: Choose a Lyapunov function of the form

$$V = \frac{1}{2} e^T e + \frac{1}{2}(\bar{\alpha} - \hat{\alpha})^T (\bar{\alpha} - \hat{\alpha}) + \frac{1}{2}(\bar{\beta} - \hat{\beta})^T (\bar{\beta} - \hat{\beta}).$$

Consequently, the time-derivative of $V$ along the error system (5) is

$$\dot{V} = e^T \dot{e} - (\bar{\alpha} - \hat{\alpha})^T \dot{\hat{\alpha}} - (\bar{\beta} - \hat{\beta})^T \dot{\hat{\beta}}$$
$$= e^T (\bar{f}(x) - \bar{g}(y) + \bar{F}(x)\bar{\alpha} - \bar{G}(y)\bar{\beta} + d(x,t) - u(t,x)) - (\bar{\alpha} - \hat{\alpha})^T \dot{\hat{\alpha}} - (\bar{\beta} - \hat{\beta})^T \dot{\hat{\beta}}$$

From above hypothesizes H1~H3 and Eq.(7), one can obtain

$$\dot{V} = e^T (\bar{f}(x) - \bar{g}(x)) + e^T (\bar{g}(x) - \bar{g}(y)) + e^T \bar{F}(x)(\bar{\alpha} - \hat{\alpha}) - e^T \bar{G}(y)(\bar{\beta} - \hat{\beta})$$
$$+ e^T d(x,t) - e^T [k_1 e + (k_2 + 1)\frac{e}{\|e\|} + 2\frac{\|\Delta_\alpha\|^2 + \|\Delta_\alpha\| \cdot \|\hat{\alpha}\|}{|\alpha_{m+1} - \hat{\alpha}_{m+1}|} \cdot \frac{e}{\|e\|^2} + 2\frac{\|\Delta_\beta\|^2 + \|\Delta_\beta\| \cdot \|\hat{\beta}\|}{|\beta_{l+1} - \hat{\beta}_{l+1}|} \cdot \frac{e}{\|e\|^2}]$$
$$- (\bar{\alpha} - \hat{\alpha})^T \dot{\hat{\alpha}} - (\bar{\beta} - \hat{\beta})^T \dot{\hat{\beta}}$$
$$\le M\|e\| + L_{\bar{g}}\|e\|^2 + e^T \bar{F}(x)(\bar{\alpha} - \hat{\alpha}) - e^T \bar{G}(y)(\bar{\beta} - \hat{\beta}) + \delta_d \|e\|$$
$$- e^T [k_1 e + (k_2 + 1)\frac{e}{\|e\|} + 2\frac{\|\Delta_\alpha\|^2 + \|\Delta_\alpha\| \cdot \|\hat{\alpha}\|}{|\alpha_{m+1} - \hat{\alpha}_{m+1}|} \cdot \frac{e}{\|e\|^2} + 2\frac{\|\Delta_\beta\|^2 + \|\Delta_\beta\| \cdot \|\hat{\beta}\|}{|\beta_{l+1} - \hat{\beta}_{l+1}|} \cdot \frac{e}{\|e\|^2}]$$
$$- (\bar{\alpha} - \hat{\alpha})^T \dot{\hat{\alpha}} - (\bar{\beta} - \hat{\beta})^T \dot{\hat{\beta}}$$

By using $k_1 \ge L_{\bar{g}}, k_2 \ge \delta_d + M$ and Eqs.( 8) and (9), it yields that

$$\dot{V} \le -\|e\| - e^T [2\frac{\|\Delta_\alpha\|^2 + \|\Delta_\alpha\| \cdot \|\hat{\alpha}\|}{|\alpha_{m+1} - \hat{\alpha}_{m+1}|} \cdot \frac{e}{\|e\|^2} + 2\frac{\|\Delta_\beta\|^2 + \|\Delta_\beta\| \cdot \|\hat{\beta}\|}{|\beta_{l+1} - \hat{\beta}_{l+1}|} \cdot \frac{e}{\|e\|^2}]$$
$$- (\bar{\alpha} - \hat{\alpha})^T \frac{\Delta_\alpha - \hat{\alpha}}{|\alpha_{m+1} - \hat{\alpha}_{m+1}|} - (\bar{\beta} - \hat{\beta})^T \frac{\Delta_\beta - \hat{\beta}}{|\beta_{l+1} - \hat{\beta}_{l+1}|}. \tag{11}$$

Using the inequalities

$$|\alpha_{m+1} - \hat{\alpha}_{m+1}| \le \|\bar{\alpha} - \hat{\alpha}\| \text{ and } |\beta_{l+1} - \hat{\beta}_{l+1}| \le \|\bar{\beta} - \hat{\beta}\|,$$

we have

$$-\frac{1}{|\alpha_{m+1}-\hat{\alpha}_{m+1}|}\le-\frac{1}{\|\bar{\alpha}-\hat{\alpha}\|} \quad \text{and} \quad -\frac{1}{|\beta_{l+1}-\hat{\beta}_{l+1}|}\le-\frac{1}{\|\bar{\beta}-\hat{\beta}\|}. \tag{12}$$

With the help of inequality (12), inequality (11) can be transformed into

$$\begin{aligned}\dot{V}\le&-\|e\|-(\bar{\alpha}-\hat{\alpha})^T\cdot\frac{\bar{\alpha}-\hat{\alpha}}{\|\bar{\alpha}-\hat{\alpha}\|}+(\bar{\alpha}-\hat{\alpha})^T\cdot\frac{\bar{\alpha}-\Delta_\alpha}{|\alpha_{m+1}-\hat{\alpha}_{m+1}|}-(\bar{\beta}-\hat{\beta})^T\cdot\frac{\bar{\beta}-\hat{\beta}}{\|\bar{\beta}-\hat{\beta}\|}\\&+(\bar{\beta}-\hat{\beta})^T\cdot\frac{\bar{\beta}-\Delta_\beta}{|\beta_{l+1}-\hat{\beta}_{l+1}|}-e^T\Big(2\frac{\|\Delta_\alpha\|^2+\|\Delta_\alpha\|\cdot\|\hat{\alpha}\|}{|\alpha_{m+1}-\hat{\alpha}_{m+1}|}\cdot\frac{e}{\|e\|^2}+2\frac{\|\Delta_\beta\|^2+\|\Delta_\beta\|\cdot\|\hat{\beta}\|}{|\beta_{l+1}-\hat{\beta}_{l+1}|}\cdot\frac{e}{\|e\|^2}\Big)\end{aligned} \tag{13}$$

The following expressions hold,

$$(\bar{\alpha}-\hat{\alpha})^T\cdot(\bar{\alpha}-\Delta_\alpha)=\|\bar{\alpha}\|^2-\bar{\alpha}^T\Delta_\alpha-\hat{\alpha}^T\bar{\alpha}+\hat{\alpha}^T\Delta_\alpha\le2\|\Delta_\alpha\|^2+2\|\Delta_\alpha\|\cdot\|\hat{\alpha}\|,$$

$$(\bar{\beta}-\hat{\beta})^T\cdot(\bar{\beta}-\Delta_\beta)=\|\bar{\beta}\|^2-\bar{\beta}^T\Delta_\beta-\hat{\beta}^T\bar{\beta}+\hat{\beta}^T\Delta_\beta\le2\|\Delta_\beta\|^2+2\|\Delta_\beta\|\cdot\|\hat{\beta}\|.$$

This leads to

$$\dot{V}\le-\|e\|-\|\bar{\alpha}-\hat{\alpha}\|-\|\bar{\beta}-\hat{\beta}\|.$$

From Lemma 2, we obtain

$$\begin{aligned}\dot{V}&\le-(\|e\|^2+\|\bar{\alpha}-\hat{\alpha}\|^2+\|\bar{\beta}-\hat{\beta}\|^2)^{\frac{1}{2}}\\&=-2^{\frac{1}{2}}V^{\frac{1}{2}}.\end{aligned}$$

By applying Lemma 1, it can be concluded that the error system (5) is stabilized at origin point in finite time $t_1$ defined by

$$t_1=t_0+\frac{V^{1-\frac{1}{2}}(t_0)}{2^{\frac{1}{2}}(1-\frac{1}{2})}=t_0+2^{\frac{1}{2}}V^{\frac{1}{2}}(t_0).$$

That is, the response system (4) synchronizes the drive system (2) in finite time.

**Remark 2.** We try to construct a suitable controller such that the specific Lyapunov function $V$ can satisfy Lemma 1 which guarantees the finite-time synchronization. The unknown parameter vectors $\alpha$ and $\beta$ in $V$ should not appear in controller and parameter update laws in practice. Using the method, for example, in [25], the controllers contain the unknown parameters. If we use the same method as above but do not use the virtual unknown parameters, the controller and parameter update laws, i.e. Eq.(7)-(9), should be

$$u(x,t) = k_1 e + (k_2+1)\frac{e}{\|e\|} + F(x)\hat{\alpha} - G(y)\hat{\beta}$$

$$+2\frac{\|\Delta_\alpha\|^2 + \|\Delta_\alpha\|\cdot\|\hat{\alpha}\|}{|\alpha_i - \hat{\alpha}_i|}\cdot\frac{e}{\|e\|^2} + 2\frac{\|\Delta_\beta\|^2 + \|\Delta_\beta\|\cdot\|\hat{\beta}\|}{|\beta_i - \hat{\beta}_i|}\cdot\frac{e}{\|e\|^2},$$

$$\dot{\hat{\alpha}} = F(x)^T e + \frac{\Delta_\alpha - \hat{\alpha}}{|\alpha_i - \hat{\alpha}_i|}, \quad \dot{\hat{\beta}} = -G(y)^T e + \frac{\Delta_\beta - \hat{\beta}}{|\beta_i - \hat{\beta}_i|},$$

which contain unknown parameters, namely $\alpha_i$ and $\beta_j$, of the drive and response systems, respectively. Theorem 1 indicates that introduction of a virtual unknown parameter is available to avoid the unknown parameters from appearing in controllers and parameters update laws and realize the finite-time synchronization of two different chaotic systems.

**Remark 3.** Obviously, the magnitude of $\frac{e}{\|e\|^2}$ in the controller $u(x,t)$ will increase unboundly and become infinity when $e \to 0$. So, we take $\frac{e}{\|e\|^2 + \varepsilon}$ instead of $\frac{e}{\|e\|^2}$ in practice, where $\varepsilon$ is a sufficient small positive constant. This idea also can be found in [26].

### 3.2 Multiple virtual unknown parameters in response system

In this part, we consider the finite time synchronization of two different chaotic systems with multiple virtual unknown parameters in response system. Sometimes it is hard to determine the Lipschitz constant. For example, if we take the Lorenz system presented in Section 2 as a response system, we have

$$\|\bar{f}(x) - \bar{f}(y)\| = \left\| \begin{pmatrix} 0 \\ -x_1x_3 + y_1y_3 \\ x_1x_2 - y_1y_2 \end{pmatrix} \right\| = \sqrt{(-x_1x_3 + y_1y_3)^2 + (x_1x_2 - y_1y_2)^2}.$$

Apparently, it is difficult to find a constant $L$ satisfying $\|\bar{f}(x) - \bar{f}(y)\| \le L\|x - y\|$. Multiple virtual unknown parameters are introduced to overcome this difficulty. The response system with controller $u(t,x)$ is chosen as

$$\dot{y}=\bar{h}(y)+\bar{H}(y)\bar{\gamma}+u(t,x)\,, \tag{14}$$

where $y\in R^n$ is the state vector, $\bar{h}(y)\in R^n$ does only contain linear terms and $\bar{H}(y)\in R^{n\times(l+s)}$ is a nonlinear matrix function. $\bar{\gamma}^T=(\gamma^T,\gamma_{l+1}\cdots\gamma_{l+s})\in R^{l+s}$ $(s\le n)$ is a new parameter vector, in which $\gamma\in R^l$ is an unknown parameter vector of the initial system and $\gamma_{l+1}\cdots\gamma_{l+s}$ are virtual unknown parameters. $u(t,x)$ is a controller. The hypotheses H1 and H2 in Theorem 1 are modified as,

**H1′.** The nonlinear functions $\bar{f}(\cdot)$ and $\bar{h}(\cdot)$ are smooth and continuous on a bounded and closed region. Thus there exists a constant $M_1$ such that

$$\|\bar{f}(x)-\bar{h}(x)\|\le M_1.$$

**H2′.** The nonlinear matrix function $\bar{h}(\cdot)$ satisfies Lipschitz condition, namely

$$\|\bar{h}(x)-\bar{h}(y)\|\le L_{\bar{h}}\|x-y\|,$$

where $L_{\bar{h}}$ is an appropriate positive constant. Then the following result is attained.

**Theorem 2.** The coupled system (2) and (14) can be synchronized in finite time if the hypotheses **H1′**, **H2′** and **H3** hold and the following conditions $(\mathrm{I})-(\Pi)$ are satisfied,

(I) The controller is chosen as

$$\begin{aligned} u(x,t)=&\,k_1e+(k_2+1)\frac{e}{\|e\|}+\bar{F}(x)\hat{\alpha}-\bar{H}(y)\hat{\gamma}\\ &+2\frac{\|\Delta_\alpha\|^2+\|\Delta_\alpha\|\cdot\|\hat{\alpha}\|}{|\alpha_{m+1}-\hat{\alpha}_{m+1}|}\cdot\frac{e}{\|e\|^2}+2\frac{\|\Delta_\gamma\|^2+\|\Delta_\gamma\|\cdot\|\hat{\gamma}\|}{|\gamma_{l+i}-\hat{\gamma}_{l+i}|}\cdot\frac{e}{\|e\|^2}. \qquad (1\le i\le s)\end{aligned} \tag{15}$$

where $k_1\ge L_{\bar{h}}$, $k_2\ge\delta_d+M_1$, $\hat{\alpha}$ and $\hat{\gamma}$ are two vector parameters, and $\Delta_\gamma=(\bar{\delta}_\gamma,\bar{\delta}_\gamma,\cdots\bar{\delta}_\gamma)^T$ is a $(l+s)$-dimensional constant vector, where $\|\bar{\gamma}\|=\bar{\delta}_\gamma$ $\le\delta_\gamma+|\gamma_{l+1}|+\cdots+|\gamma_{l+s}|,1\le s\le n$.

(Π) The update laws of the parameters $\hat{\alpha}$ and $\hat{\gamma}$ are

$$\dot{\hat{\alpha}} = \bar{F}(x)^T e + \frac{\Delta_\alpha - \hat{\alpha}}{|\alpha_{m+1} - \hat{\alpha}_{m+1}|}, \tag{16}$$

$$\dot{\hat{\gamma}} = -\bar{H}(z)^T e + \frac{\Delta_\gamma - \hat{\gamma}}{|\gamma_{l+i} - \hat{\gamma}_{l+i}|}. \qquad (1 \le i \le s) \tag{17}$$

where $\hat{\alpha}_{m+1}$ and $\hat{\gamma}_{l+i}$ represent the $(m+1)-th$ and $(l+i)-th$ component of vectors $\hat{\alpha}$ and $\hat{\gamma}$, respectively.

Furthermore, for any given $t_0$, the finite time $t_1$ is determined by $t_1 = t_0 + 2^{\frac{1}{2}} V^{\frac{1}{2}}(t_0)$, where $V$ is a positive-definite function satisfying Lemma 1.

So we can draw a conclusion that the response system (14) can globally synchronize the drive system (2) in finite time. The proof procedure of Theorem 2 is similar to that of Theorem 1 and omitted here.

## 4. Numerical simulation

To verify the feasibility of the proposed method, the Rössler system and modified Chua system, Rössler system and Lorenz system are used as examples in the simulations.

**Example 1** The Rössler system and modified Chua system for the case of Theorem 1

The drive system is Rössler system, with disturbance $d(x,t)$ and unknown parameter $\alpha$ described by

$$\begin{aligned}\begin{pmatrix}\dot{x}_1\\ \dot{x}_2\\ \dot{x}_3\end{pmatrix} &= \begin{pmatrix}0\\ x_1\\ x_1x_3+0.2\end{pmatrix} + \begin{pmatrix}x_2+x_3 & 0 & 0\\ 0 & x_2 & 0\\ 0 & 0 & x_3\end{pmatrix}\begin{pmatrix}\alpha_1\\ \alpha_2\\ \alpha_3\end{pmatrix} + \begin{pmatrix}d_1(x,t)\\ d_2(x,t)\\ d_3(x,t)\end{pmatrix}\\ &= \begin{pmatrix}0\\ 0\\ x_1x_3+0.2\end{pmatrix} + \begin{pmatrix}x_2+x_3 & 0 & 0 & 0\\ 0 & x_2 & 0 & x_1\\ 0 & 0 & x_3 & 0\end{pmatrix}\begin{pmatrix}\alpha_1\\ \alpha_2\\ \alpha_3\\ \alpha_4\end{pmatrix} + \begin{pmatrix}d_1(x,t)\\ d_2(x,t)\\ d_3(x,t)\end{pmatrix}\end{aligned} \tag{18}$$

The controlled modified Chua' circuit is used for the response system[27]

$$
\begin{pmatrix} \dot{y}_1 \\ \dot{y}_2 \\ \dot{y}_3 \end{pmatrix} = \begin{pmatrix} 0 \\ y_1 - y_2 + y_3 \\ 0 \end{pmatrix} + \begin{pmatrix} y_2 - \frac{1}{7}(2y_1^3 - y_1) & 0 \\ 0 & 0 \\ 0 & -y_2 \end{pmatrix} \begin{pmatrix} \beta_1 \\ \beta_2 \end{pmatrix} + \begin{pmatrix} u_1(x,t) \\ u_2(x,t) \\ u_3(x,t) \end{pmatrix}
$$

$$
= \begin{pmatrix} 0 \\ y_1 - y_2 \\ 0 \end{pmatrix} + \begin{pmatrix} y_2 - \frac{1}{7}(2y_1^3 - y_1) & 0 & 0 \\ 0 & 0 & y_3 \\ 0 & -y_2 & 0 \end{pmatrix} \begin{pmatrix} \beta_1 \\ \beta_2 \\ \beta_3 \end{pmatrix} + \begin{pmatrix} u_1(x,t) \\ u_2(x,t) \\ u_3(x,t) \end{pmatrix} \tag{19}
$$

In simulation, values of uncertain parameters of the drive and response systems are chosen as $\alpha_1 = -1, \alpha_2 = 0.2, \alpha_3 = -7, \beta_1 = 10, \beta_2 = \frac{100}{7}$. The virtual unknown parameters are $\alpha_4 = \beta_3 = 1$. The small positive constant is set to $\varepsilon = 10^{-4}$, and the disturbance is $d = (d_1, d_2, d_3)^T = (0.05x_2 \sin t, 0.1x_3 \sin 3t, 0.01x_1 \cos 2t)^T$. Figs.1-2 shows that the drive system (18) is chaotic under the above parameter values and the initial value $(x_1(0), x_2(0), x_3(0))^T = (1,1,1)^T$. From Figs.1-2, the bounds of the chaotic attractor are $-10 < x_1 < 11.5, -11 < x_2 < 8, 0 < x_3 < 24$. Therefore, the following auxiliary values can be obtained

$$
\|\delta_d\| \le 2.7, \quad \|\delta_\alpha\| \le 7.1, \|\bar{\delta}_\alpha\| \le 8.1, \quad \|\delta_\beta\| \le 17.6, \|\bar{\delta}_\beta\| \le 18.6
$$

$$
\|\bar{f}(x) - \bar{g}(x)\| = \left\| \begin{pmatrix} 0 \\ -x_1 + x_2 \\ x_1 x_3 + 0.2 \end{pmatrix} \right\| = \sqrt{(-x_1 + x_2)^2 + (x_1 x_3 + 0.2)^2} \le 278,
$$

It is easy to verify that $\bar{g}(\cdot)$ satisfies Lipschitz condition,

$$
\|\bar{g}(x) - \bar{g}(y)\| = \left\| \begin{pmatrix} 0 \\ x_1 - x_2 - y_1 + y_2 \\ 0 \end{pmatrix} \right\| \le 2\|x - y\|.
$$

By Theorem 1, we chose $k_1 = 3$ and $k_2 = 281$. The controller is given by the following expression,

$$u = 3e + \frac{281}{\|e\|}e + \bar{F}(x)\hat{\alpha} - \bar{G}(y)\hat{\beta}$$
$$+2\frac{\|\Delta_\alpha\|^2 + \|\Delta_\alpha\|\cdot\|\hat{\alpha}\|}{|1-\hat{\alpha}_4|}\cdot\frac{e}{\|e\|^2} + 2\frac{\|\Delta_\beta\|^2 + \|\Delta_\beta\|\cdot\|\hat{\beta}\|}{|1-\hat{\beta}_3|}\cdot\frac{e}{\|e\|^2},$$

where $\hat{\alpha}, \hat{\beta}$ are updated according to the following adaptation algorithm,

$$\dot{\hat{\alpha}} = \begin{pmatrix} x_2 + x_3 & 0 & 0 \\ 0 & x_2 & 0 \\ 0 & 0 & x_3 \\ 0 & x_1 & 0 \end{pmatrix} e + \frac{1}{|1-\hat{\alpha}_4|}\begin{pmatrix} 8.1 - \hat{\alpha}_1 \\ 8.1 - \hat{\alpha}_2 \\ 8.1 - \hat{\alpha}_3 \\ 8.1 - \hat{\alpha}_4 \end{pmatrix},$$

$$\dot{\hat{\beta}} = \begin{pmatrix} -y_2 + \frac{1}{7}(2y_1^3 - y_1) & 0 & 0 \\ 0 & 0 & y_2 \\ 0 & -y_3 & 0 \end{pmatrix} e + \frac{1}{|1-\hat{\beta}_3|}\begin{pmatrix} 18.6 - \hat{\beta}_1 \\ 18.6 - \hat{\beta}_2 \\ 18.6 - \hat{\beta}_3 \end{pmatrix}.$$

The initial conditions for the response system and update laws of the parameters are $(y_1(0), y_2(0), y_3(0))^T = (-1, 2, -1)^T$, $(\hat{\alpha}_1(0), \hat{\alpha}_2(0), \hat{\alpha}_3(0), \hat{\alpha}_4(0))^T = (1, 2, 3, 4)^T$, $(\hat{\beta}_1(0), \hat{\beta}_2(0), \hat{\beta}_3(0))^T = (0.5, 1, 1.5)^T$, respectively. From Fig.3, one can see the error vector $e$ converges to zero at $t = 0.002$ with control operation. This implies that the trajectories of the response system converge to those of the drive system in finite time. While by the Eq.(10), the converge time is $t = 16.8148$. The difference between the numerical and theoretical converge time mostly comes from the conservative chose of $k_1$ and $k_2$.

**Example 2** The Rössler system and Lorenz system for the case of Theorem 2

Let Eq.(17) be a drive system. The controlled Lorenz system with following equation is used for response system,

$$\begin{pmatrix} \dot{y}_1 \\ \dot{y}_2 \\ \dot{y}_3 \end{pmatrix} = \begin{pmatrix} 0 \\ -y_2 - y_1 y_3 \\ y_1 y_2 \end{pmatrix} + \begin{pmatrix} y_2 - y_1 & 0 & 0 \\ 0 & y_1 & 0 \\ 0 & 0 & y_3 \end{pmatrix}\begin{pmatrix} \gamma_1 \\ \gamma_2 \\ \gamma_3 \end{pmatrix} = h(y) + H(y)\gamma,$$

$$= \begin{pmatrix} 0 \\ -y_2 \\ 0 \end{pmatrix} + \begin{pmatrix} y_2 - y_1 & 0 & 0 & 0 & 0 \\ 0 & y_1 & 0 & -y_1 y_3 & 0 \\ 0 & 0 & y_3 & 0 & y_1 y_2 \end{pmatrix} \begin{pmatrix} \gamma_1 \\ \gamma_2 \\ \gamma_3 \\ \gamma_4 \\ \gamma_5 \end{pmatrix} = \bar{h}(y) + \bar{H}(y)\bar{\gamma}. \qquad (20)$$

Apparently, finding a Lipchitz constant for nonlinear function $h(y)$ needs the bounds of the response system, which are difficult to be determined in practice. Introduction of multiple virtual unknown parameters can overcome this difficulty. Values of uncertain parameters of the response system are chosen as $\gamma_1 = 10, \gamma_2 = \frac{8}{3}, \gamma_3 = 28$, and the virtual unknown parameters are $\gamma_4 = \gamma_5 = 1$. Therefore, the following auxiliary values can be got

$$\|\delta_\gamma\| \le 30, \|\bar{\delta}_\gamma\| \le 32, \Delta_\gamma = (32,32,32,32,32)^T.$$

$$\|\bar{f}(x) - \bar{h}(x)\| = \left\| \begin{pmatrix} 0 \\ x_2 \\ x_1 x_3 + 0.2 \end{pmatrix} \right\| = \sqrt{x_2^{\,2} + (x_1 x_3 + 0.2)^2} \le 276.5,$$

It is easy to verify that $\bar{h}(\cdot)$ satisfies Lipschitz condition,

$$\|\bar{h}(x) - \bar{h}(y)\| = \left\| \begin{pmatrix} 0 \\ -x_2 + y_2 \\ 0 \end{pmatrix} \right\| \le \|x - y\|.$$

Proceeding as before, we chose $k_1 = 2$ and $k_2 = 280$. The controller is given by the following expression,

$$u = 2e + \frac{280}{\|e\|} e + \bar{F}(x)\hat{\alpha} - \bar{H}(y)\hat{\gamma}$$
$$+2\frac{\|\Delta_\alpha\|^2 + \|\Delta_\alpha\| \cdot \|\hat{\alpha}\|}{|1-\hat{\alpha}_4|} \cdot \frac{e}{\|e\|^2} + 2\frac{\|\Delta_\gamma\|^2 + \|\Delta_\gamma\| \cdot \|\hat{\gamma}\|}{|1-\hat{\gamma}_5|} \cdot \frac{e}{\|e\|^2},$$

where $\hat{\gamma}$ is updated according to the following adaptation algorithm,

$$\dot{\hat{\gamma}}=\begin{pmatrix} y_1-y_2 & 0 & 0 \\ 0 & -y_1 & 0 \\ 0 & 0 & -y_3 \\ 0 & y_1y_3 & 0 \\ 0 & 0 & -y_1y_2 \end{pmatrix} e+\frac{1}{|1-\hat{\gamma}_5|}\begin{pmatrix} 32-\hat{\gamma}_1 \\ 32-\hat{\gamma}_2 \\ 32-\hat{\gamma}_3 \\ 32-\hat{\gamma}_4 \\ 32-\hat{\gamma}_5 \end{pmatrix}.$$

The initial condition of $\hat{\gamma}$ is $(\hat{\gamma}_1(0),\hat{\gamma}_2(0),\hat{\gamma}_3(0),\hat{\gamma}_4(0),\hat{\gamma}_5(0))^T=(5,6,7,0,1)^T$. The other values of parameters and the adaptation laws of $\hat{\alpha}$ are the same as those of Example 1. From Fig.4, one can see the error vector $e$ converges to zero at $t=0.0008$ with control operation. By the Eq.(10), the theoretical converge time is $t=29.2494$.

## 5. Conclusion

This work presents a general method for synchronizing two different chaotic systems with disturbance and unknown parameters in finite time. We introduce virtual unknown parameters to avoid the unknown parameters from appearing in controllers and parameters update laws. A single virtual unknown parameter is used to design an adaptive controller and the parameter update laws which guarantee the finite-time synchronization between two coupled chaotic systems. For the case that Lipschitz constant is hard to be determined, multiple virtual parameters are taken to develop techniques for obtaining proper adaptive controllers. Numerical simulations on the basis of the Rössler system and Chua's circuit, Rössler system and Lorenz system are presented to show the effectiveness of the proposed method.

**Acknowledgement** This work is supported by the Foundation for Supporting Universities in Fujian Province of China under grant No JK2009020.

## References

[1] S. Boccaletti, J. Kurths, G. Osipov, D.L. Valladares, C.S. Zhou. The synchronization of chaotic systems, *Physics Reports,* 2002, 366:1-101.

[2] F. Liu, Y. Ren, X.M. Shan, Z.L. Qiu. A linear feedback synchronization theorem

for a class of chaotic systems, *Chaos, Solitons and Fractals,* 2002, 13:723-730**.**

[3] C.H. Tao, H.X. Xiong, F. Hu. Two novel synchronization criterions for a unified chaos system, *Chaos, Solitons and Fractals,* 2006, 27:115-120**.**

[4] H. Zhang, X.K. Ma, W.Z. Liu. Synchronization of chaotic systems with parametric uncertainty using active sliding mode control, *Chaos, Solitons and Fractals*, 2004, 21:1249-1257.

[5] H.T. Yau. Design of adaptive sliding mode controller for chaos synchronization with uncertainties, *Chaos, Solitons and Fractals*, 2004, 22: 341-347.

[6] H. Fotsin, S. Bowong, J. Daafouz. Adaptive synchronization of two chaotic systems consisting of modified Van der Pol-Duffing and Chua oscillators, *Chaos, Solitons and Fractals,* 2005, 26:215-229.

[7] M.T. Yassen. Adaptive synchronization of two different uncertain chaotic systems, *Physics Letter A*, 2005, 337: 335-341.

[8] S.H. Li, Y.P. Tian. Finite time synchronization of chaotic systems, *Chaos, Solitons and Fractals*, 2003, 15: 303-310.

[9] Y.G. Hong. Finite-time stabilization and stabilizability of a class of controllable systems, *Systems and Control Letters*, 2002, 46: 231-236.

[10] X. Q. Huang, W. Lin, B. Yang. Global finite-time stabilization of a class of uncertain nonlinear systems, *Automatica*, 2005, 41: 881-888.

[11] J. Li and C.J. Qian. Global finite-time stabilization by dynamic output feedback for a class of continuous nonlinear systems, *IEEE Transactions on automatic control*, 2006, 51: 879-884.

[12] Y.G. Hong, H.K. Wang, D.Z. Cheng. Adaptive finite-time control of nonlinear systems with parametric uncertainty，*IEEE Transactions on automatic control*, 2006, 51: 858-862.

[13] C.J. Qian and J. Li. Global finite-time stabilization by output feedback for planar systems without observable linearization, *IEEE Transactions on automatic control*, 2005, 50:885-890

[14] Y.G. Hong, J.K. Wang. Finite time stabilization for a class of nonlinear systems,

*8th International Conference on Control, Automation, Robotics and Vision*, Kunming, 2004, pp. 1194-1199.

[15]J. Li, C.J. Qian, Global Finite-Time Stabilization of a class of uncertain nonlinear systems using output feedback, *44th IEEE Conference on Decision and Control, and the European Control Conference*, Seville, 2005, pp. 2652-2657.

[16] S.Onori, P.Dorato, S.Galeani, C.T.Abdallah. Finite time stability design via feedback linearization, *44th IEEE Conference on Decision and Control, and the European Control Conference*, Seville, 2005, pp.4915-4920.

[17] W. Perruquetti, T. Floquet, E. Moulay. Finite-time observers: Application to secure communication, *IEEE Transactions on automatic control*, 2008, 53(1):356-360.

[18] W. Perruquetti and T. Floquet. Homogeneous finite time observer for nonlinear systems with linearizable error dynamics, *46th IEEE Conference on Decision and Control*, New Orleans, 2007, pp. 390-395.

[19] Y.G. Hong, G.U. Yang, L. Bushnell, H. Wang. Global finite-time stabilization: from state feedback to output feedback, *39th IEEE Conference on Decision and Control*, Sydney, 2000, pp. 2908-2913.

[20] F. Amato, M. Ariola, P. Dorato. Finite-time control of linear systems subject to parametric uncertainties and disturbances, *Automatica,* 2001, 37: 1459-1463.

[21]T.G. Gao, Z.Q. Chen, G.R. Chen, Z.Z. Yuan. Finite-time control of chaotic systems with nonlinear imputs, *Chinese Physics*, 2006, 15: 1190-1195.

[22]H. Wang, Z.Z. Han, Q.Y. Xie, W. Zhang. Finite-time chaos control via nonsingular terminal sliding mode control, *Commun Nonlinear Sci Numer Simulat*，2009, 14: 2728-2733.

[23] H. Wang, Z.Z. Han, Q.Y. Xie, W. Zhang. Finite-time chaos synchronization of unified chaotic system with uncertain parameters, *Commun Nonlinear Sci Numer Simulat*, 2009, 14: 2239-2247.

[24] H. Wang, Z.Z. Han, Q.Y. Xie, W. Zhang. Finite-time chaos synchronization of unified chaotic systems base on CLF, *Nonlinear Analysis: Real World Applications*, 2009, 10: 2842-2849.

[25] Z.E. Ge, J.W. Chen. Chaos synchronization and parameter identification of three time scales brushless DC motor system, *Chaos, Solitons and Fractals*, 2005, 24: 597-616.

[26] X.C. Li, W. Xu, Y.Z. Xiao. Adaptive tracking control of a class of uncertain chaotic systems in the presence of random perturbations, *Journal of Sound and Vibration,* 2008, 314: 526-535.

[27]J.J. Yan, J.S. Lin, T.L. Liao. Synchronization of a modified Chua's circuit system via adaptive sliding mode control, *Chaos, Solitons and Fractals*, 2008, 36: 45-52.

**Figure captions**

Fig.1 $x_1 - x_2$ plane projection of Rössler attractor

Fig.2 $x_1 - x_3$ plane projection of Rössler attractor

Fig.3 Synchronization error between systems (18) and (19) with initial conditions $(x_1, x_2, x_3) = (1,1,1)$ and $(y_1, y_2, y_3) = (-1,2,-1)$.

Fig.4 Synchronization error between systems (18) and (20) with initial conditions $(x_1, x_2, x_3) = (1,1,1)$ and $(y_1, y_2, y_3) = (-1,2,-1)$.

## Figures

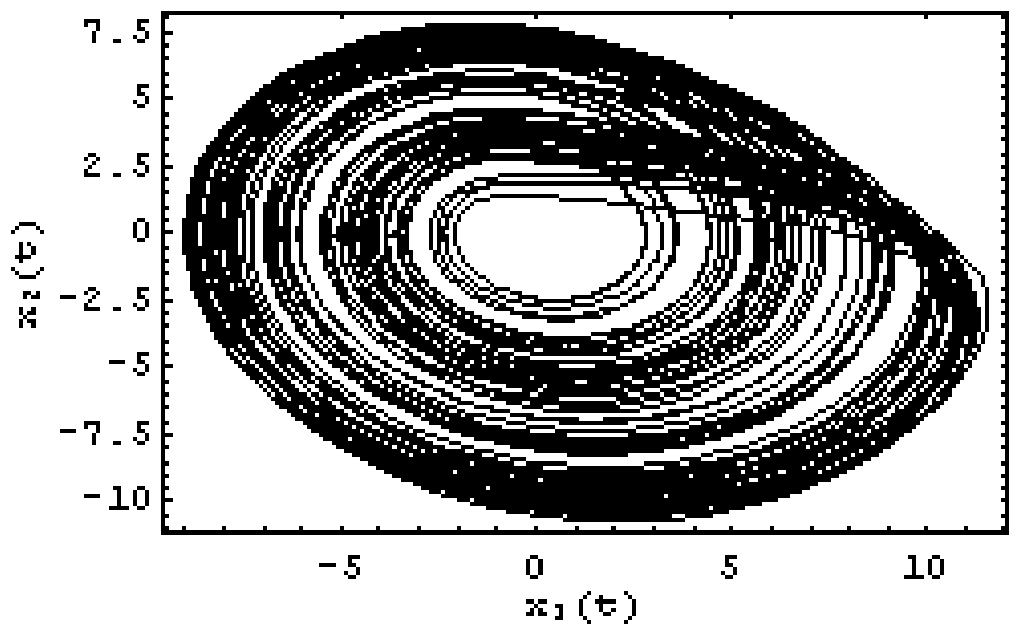

Fig.1 $x_1 - x_2$ plane projection of Rössler attractor

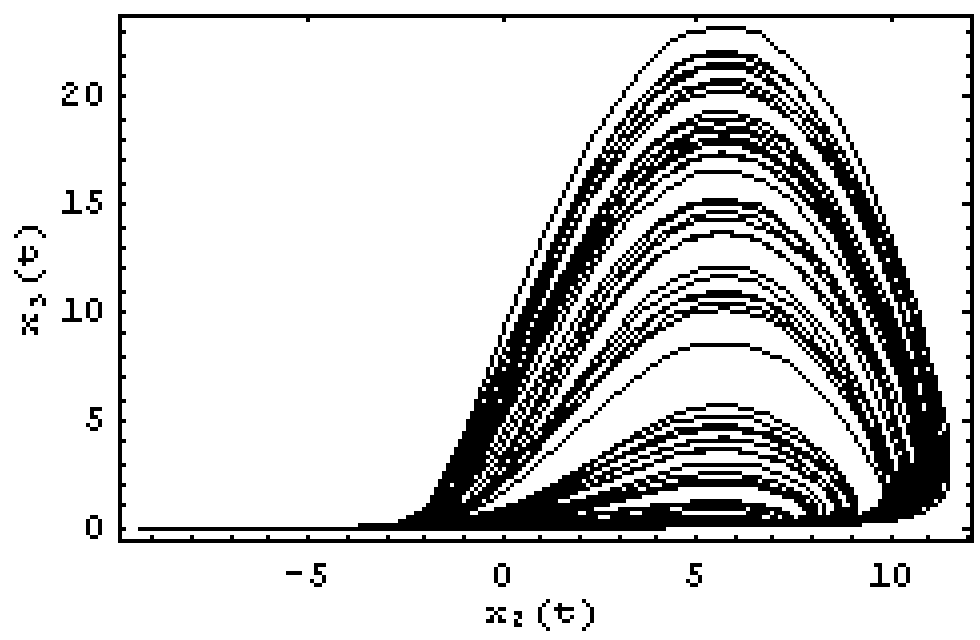

Fig.2 $x_1 - x_3$ plane projection of Rössler attractor

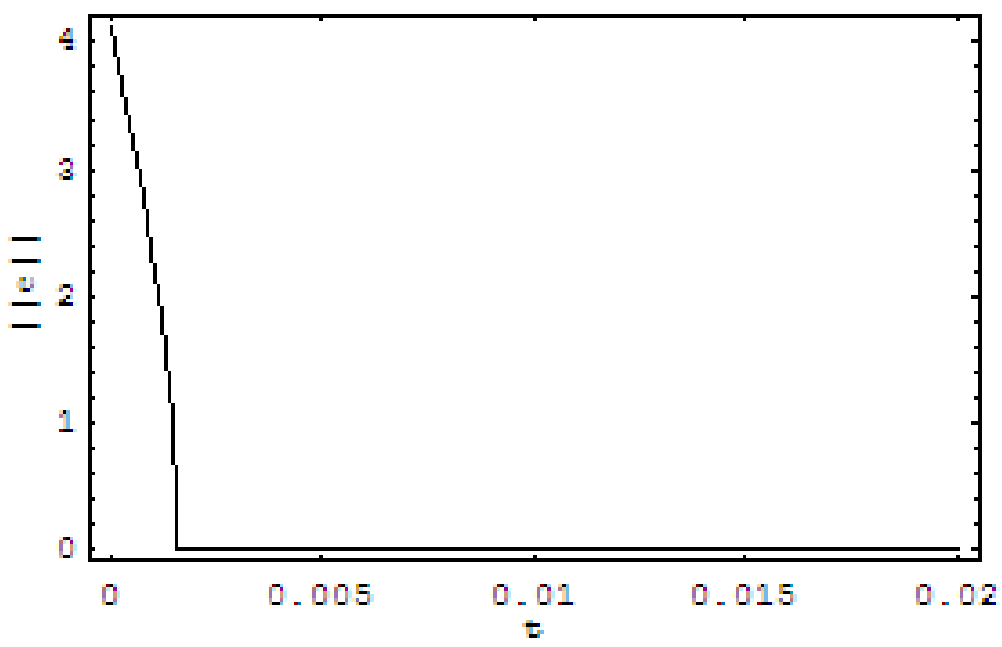

Fig.3 Synchronization error between systems (18) and (19) with initial conditions

$(x_1, x_2, x_3) = (1,1,1)$ and $(y_1, y_2, y_3) = (-1,2,-1)$.

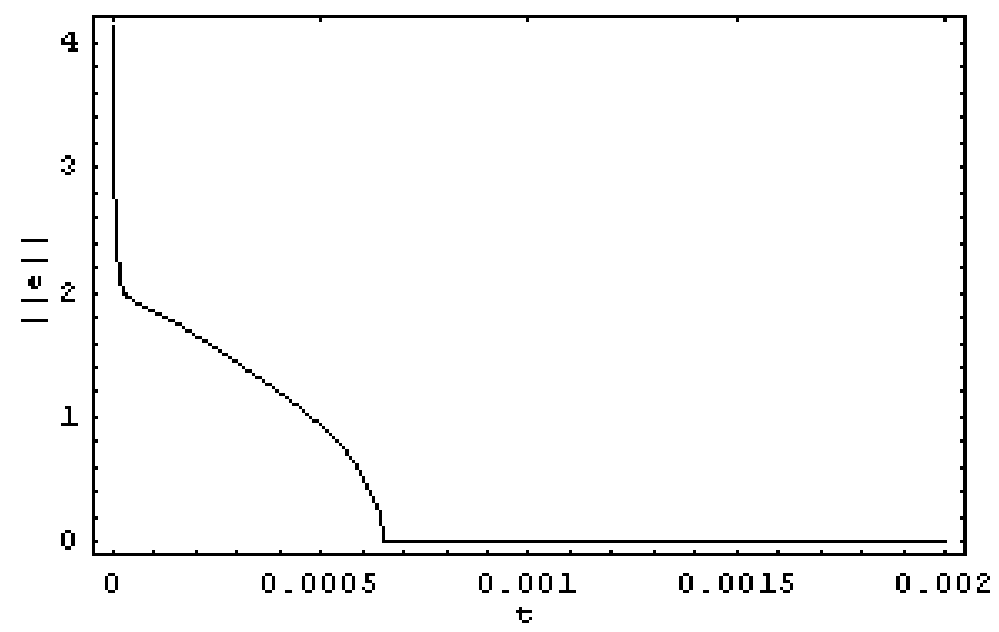

Fig.4 Synchronization error between systems (18) and (20) with initial conditions

$(x_1, x_2, x_3) = (1,1,1)$ and $(y_1, y_2, y_3) = (-1,2,-1)$.